\documentclass[twocolumn,showpacs,prc,aps,floatfix]{revtex4-1}

\usepackage{amssymb}
\usepackage{graphicx}
\usepackage{dcolumn}
\usepackage{amsmath}
\usepackage{amsfonts}
\usepackage{amssymb}
\usepackage{xcolor}
\usepackage{braket}
\usepackage{relsize}
\usepackage{float} 
\usepackage{hyperref}
\usepackage{bm}
\usepackage{mathtools}
\mathtoolsset{showonlyrefs}

\hypersetup{
    colorlinks,
    linkcolor={blue},
    citecolor={blue},
    urlcolor={blue}
}

\begin{document}

\title{Reexamining the relation between the binding energy of finite nuclei and the equation of state of infinite nuclear matter}
\author{M. C. Atkinson$^{1,2}$, W. H. Dickhoff$^1$, M. Piarulli$^1$, A. Rios$^3$, and R. B. Wiringa$^4$} 
\affiliation{${}^1$Department of Physics,
Washington University, St. Louis, Missouri 63130, USA}
\affiliation{${}^2$TRIUMF, Vancouver, British Columbia V6T 2A3, Canada}
\affiliation{${}^3$Department of Physics, Faculty of Engineering and Physical Sciences, University of Surrey, Guildford, Surrey GU2 7XH, United Kingdom}
\affiliation{${}^4$Physics Division, Argonne National Laboratory, Argonne, Illinois 60439, USA}

\date{\today}

\begin{abstract}
   The energy density is calculated in coordinate space for $^{12}$C, $^{40}$Ca, $^{48}$Ca, and
   $^{208}$Pb using a dispersive optical model constrained by all relevant data including the
   corresponding energy of the ground state.  
   The energy density of $^{8}$Be is also calculated using the Green's function Monte-Carlo method
   employing the Argonne/Urbana two and three-body interactions.  The nuclear interior minimally
   contributes to the total binding energy due to the 4$\pi r^2$ phase space factor.  Thus, the
   volume contribution to the energy in the interior is not well constrained.  The dispersive-optical-model energy
   densities are in good agreement with \textit{ab initio} self-consistent Green's function
   calculations of infinite nuclear matter restricted to treat only short-range and tensor
   correlations.  These results call into question the degree to which the equation of state for
   nuclear matter is constrained by the empirical mass formula. In particular, the results in this
   paper indicate that saturated nuclear matter does not require the canonical value of 16 MeV
   binding per particle but only about 13-14 MeV when the interior of $^{208}$Pb is considered. 
\end{abstract}

\maketitle

\section{Introduction}
\label{sec:intro}
The investigation of the binding energy of atomic nuclei dates back to the origins of nuclear
physics~\cite{Bohr-Mottelson}. 
The well-known empirical mass formula, developed by Bethe and Weizs\"acker~\cite{Bethe:1936,Weizsacker:1935}, accurately describes the global aspects of nuclear binding for most of the nuclear chart. 
Its success is largely due to the saturating nature of the constituent nucleons in
nuclei. The evidence for nuclear saturation came from measurements of the root-mean-squared (rms) charge radius of
nuclei which revealed that the volume of a given nucleus scales linearly with $A$~\cite{Bethe:1971,Bohr-Mottelson}.
Elastic electron-scattering experiments revealed that the density in the interior of nuclei saturates at a value around
$\rho_0 \approx 0.16$ fm$^{-3}$~\cite{Hofstadter:1957,Bethe:1971}. In order to understand the mechanism behind nuclear
saturation, infinite nuclear matter (NM) is an ideal system that is often
studied~\cite{Exposed!,Brueckner:1954,Bethe:1956}.  
Depending on the method and realistic nucleon-nucleon (NN) interaction used, the calculated value of $\rho_0$ in NM can stray from the experimental value as discussed \textit{e.g.} in Ref.~\cite{Baldo:2012}.  
In addition to the density at saturation, the associated binding energy, $E_0$, plays a vital role in the equation of
state (EOS) of NM. The EOS does not exhibit saturation in neutron-rich systems, but its characterization is nonetheless relevant for astrophysical research on supernovae and neutron stars~\cite{Horowitz01,Steiner10,APR:1998}.

The traditional method used to estimate $\rho_0$ is fundamentally different than that of $E_0$.
While the value of $\rho_0$ is determined experimentally,  $E_0$ is determined empirically from an extrapolation of the empirical mass formula~\cite{Myers:1996,Jeukenne:1976,Bethe:1971}
\begin{align}
   BE(A,Z) = -a_V A + a_S A^{2/3} &+ a_CZ(Z-1)A^{-1/3} \nonumber\\
   + \frac{1}{2}&a_A(A-2Z)^2A^{-1},
   \label{eq:mass_formula}
\end{align}
where $a_V$, $a_S$, $a_C$, and $a_A$ are parameters fit to nuclear masses~\cite{Bohr-Mottelson}.
Because the only link between Eq.~\eqref{eq:mass_formula} and NM is the volume term, the canonical value of the saturation
energy is assumed to be $E_0/A = -a_V \approx -16$~MeV~\cite{Bethe:1971,Myers:1996}.
However, this involves a significant extrapolation that neglects proper consideration of long-range
correlations (LRC) in both
finite and infinite systems~\cite{Dewulf:2003,Dickhoff04,Exposed!,Dickhoff:2016}.
Contributions to the binding energy from LRC are associated with collective phenomena. 
In finite nuclei, these emerge as low-lying natural parity surface vibrations and higher-lying giant resonances.
These excitations are associated with the presence of a surface and therefore have no counterpart in NM.
Conversely, LRC in NM are characterized by their total momentum (and spin-isospin quantum numbers) which have no direct counterpart in finite nuclei as momentum is not a good quantum number of an excited state in a nucleus.
This is particularly problematic for matter excitations with pionic quantum numbers as the related soft mode in NM occurs at finite momentum and thereby contributes substantially to binding, is strongly enhanced by the coupling to the $\Delta$-isobar, and increases in importance with density. 
For this reason, it was argued in Ref.~\cite{Dewulf:2003} that the link between finite nuclei and NM saturation properties should be confined to the effect of short-range correlations (SRC).
Assumptions made about the role of LRC therefore influence the link between finite nuclei and NM.
As will be shown below, it is possible to establish such a link using the \textit{ab initio} method of self-consistent Green's functions (SCGF).  
We therefore propose to exercise caution when equating a fundamental property of NM to a parameter that relies heavily on the chosen functional form of the empirical mass formula. 

The mass formula of Eq.~\eqref{eq:mass_formula} is built upon the liquid drop model (LDM) of finite nuclei. The LDM has been studied and modified several times. 
These modifications mainly involved accounting for deformation, shell effects, and pairing. 
A recent form of the LDM, known as the finite-range droplet model, has improved
agreement with experimental masses~\cite{Lunney:2003}. Additionally, there have
been many other macroscopic mass models such as that of
Duflo-Zuker~\cite{Duflo:1995}, Koura~\cite{Koura:2000}, and others (see
Ref.~\cite{Lunney:2003} for a review of mass models). The parameters of
Eq.~\eqref{eq:mass_formula} have also been analyzed using different methods of
statistical analysis resulting in errors in the range of 0.03 - 0.24 for
$a_V$~\cite{Bertsch:2017}.  Much work has been focused on the parameters of
Eq.~\eqref{eq:mass_formula}, but not the connection between $a_V$ and $E_0$.
While all nuclear mass models like Eq.~\eqref{eq:mass_formula} show good
agreement with experimental masses, none address the issue of the contribution
of LRC as discussed above.

To further explore the extrapolation from the LDM to NM, consider the analogous
infinite system of liquid Helium. 
Quantum Monte Carlo studies of drops of atomic Helium,
both bosonic $^4$He~\cite{PZPWH83} and fermionic $^3$He~\cite{PPW86} using the HFDHE2 atom-atom interaction~\cite{Aziz79}, are able
to extract a reasonable volume binding energy from finite drops in a liquid
drop mass formula only by including additional terms beyond the standard
volume and surface terms of Eq.~(\ref{eq:mass_formula}). 
For the $^3$He case, fitting the energies of drops containing up to 240 atoms with
only
volume and surface
terms predicts a volume binding energy of $-1.42$ K while
adding a curvature term 
$\propto A^{1/3}$ 
generates a much better fit with a
volume term of $-2.09$ K. This is much closer to the infinite liquid result of
$-2.36$ K and the experimental value of $-2.47$ K.  
The extrapolated energy of the infinite system is
highly dependent on the chosen functional form of the LDM. The discrepancy
between the experimental binding energy and the volume energy of the LDM for
liquid $^3$He 
indicates that the traditional extrapolation to an infinite system is
insufficient even for a system with only a simple central interaction. 

An alternative connection between the physics of finite nuclei and that of NM is provided by energy density functionals (EDFs) used in nuclear density functional theories (DFTs). The EDF provides a one-to-one correspondence between binding energy and density based on effective forces such as Skyrme or Gogny. These EDFs are parametrized by fits and used to self-consistently solve for the ground state density of nuclei with Kohn-Sham (or Hartree-Fock) type equations~\cite{Bender:2003}. 
A result of these calculations is a nucleus-dependent energy density profile which is used to calculate the total binding energy. These more microscopic approaches are very successful in calculating
   binding energies and other properties across the nuclear chart~\cite{Bender:2003}. 
   The value of $E_0$ can be calculated directly from the EDF parameters. However, in the vast majority of Skyrme and Gogny EDFs, $E_0$ is a parameter of the fit rather than a prediction (or extrapolation) from properties of finite nuclei~\cite{Brack:1985,Dutra:2012,Kortelainen:2010}. Alternatively, some EDFs, such as the so-called SV-min, fit to $a_V$ in a $\chi^2$ minimization procedure resulting in statistical uncertainties around $0.06$ MeV for $E_0$~\cite{Klupfel:2009}. Other systematic studies of Skyrme EDFs reveal a similar range of $E_0$ values that allow for an acceptable reproduction of finite-nucleus data~\cite{Reinhard:2006,Klupfel:2009,Margueron:2018}.
While the fact that these EDFs can simultaneously reproduce nuclear masses and $E_0\approx-16$ MeV supports the canonical value, so far EDF calculations do not provide an extraction of $E_0$ from finite nuclear data where systematics have been explored in detail. 

In the present paper we discuss various ingredients that address some of the issues related to determining the
saturation point of symmetric NM and re-examine the empirical value of $E_0$.  This is done by comparing three different
methods of obtaining the value of $E_0$: 
the canonical value obtained from an empirical mass formula extrapolation ($a_V$), 
the minimum energy in NM from \textit{ab initio} SCGF simulations, 
and the energy density in the interior of finite nuclei based on the dispersive optical model (DOM). 
In Sec.~\ref{sec:DOM},
we present results from DOM calculations of several nuclei that are constrained, in addition to scattering observables, by ground-state properties including the energy.  By casting these results in terms of an energy density, we show in Sec.~\ref{sec:abin} that it is possible to make contact with \textit{ab initio} SCGF calculations of symmetric and asymmetric NM~\cite{Baldo:2012,Rios:2014}. 
The DOM ground-state energy is calculated using the Migdal-Galitski sum rule, so it does not explicitly include three-body forces~\cite{Galitski:1958}. 
We address this issue by utilizing energy densities from Monte Carlo calculations obtained using
various chiral two- and three-body interactions~\cite{Piarulli:2016,Piarulli:2018,Baroni:2018} as well as the phenomenological Argonne/Urbana combination~\cite{Pudliner:1995}. Further
analysis of DOM nuclear energy densities is presented in Sec.~\ref{sec:analysis} before the conclusions in Sec.~\ref{sec:conclusions}.

\section{Dispersive optical model approach}
\label{sec:DOM}
Ideally, a sound determination of $E_0$ would rely on a NM theoretical calculation based on the true NN interaction (\textit{i.e.}~obtained as a solution of the quantum chromodynamics Lagrangian). In practice, calculations of the saturation point of NM are hampered by approximations in the NN forces, limited by the treatment of three-nucleon (NNN) interactions, and display a substantial dependence on the employed many-body method~\cite{Baldo:2012,Hagen:2014,Lonardoni:2019,Drischler:2019}. This scheme dependence 
ultimately undermines a direct, reliable determination of $E_0$. 
   Even so, recent works have explored the link between finite nuclei and NM using coupled cluster and many-body perturbation theory starting from chiral NN+NNN interactions~\cite{Drischler:2020,Jiang:2020,Sammarruca:2020}. By including the empirical saturation point in their fits, low-energy constants (LECs) associated with the NNN interaction can be adjusted to reproduce the empirical saturation point~\cite{Sammarruca:2020,Jiang:2020,Drischler:2019}. However, this adjustment to the LECs leads to under-binding in finite nuclei~\cite{Sammarruca:2020}, demonstrating the difficulty in simultaneously reproducing finite nuclear binding energies and the empirical saturation point from chiral interactions. 
   Moreover, recent advances in quantifying theoretical uncertainties in NM calculations can be found in Ref.~\cite{Drischler:2020}. While the interactions used were tailored to reproduce the empirical saturation point, the newly developed Bayesian machine-learning method provides a step forward in NM calculations from chiral interactions.

Alternatively, we investigate the connection
between the empirical mass formula and the value of $E_0$ through energy
densities calculated using the DOM. This method constrains a complex self-energy  $\Sigma_{\ell j}$ using both scattering and bound-state data~\cite{Mahaux91,Mahzoon:2014}. 
The self-energy is a complex, nonlocal, energy-dependent potential that unites the nuclear structure and reaction domains through dispersion relations~\cite{Mahaux91,Mahzoon:2014,Dickhoff:2017}.
The Dyson equation generates the single-particle propagator, or Green's function, $G_{\ell j}(r,r';E)$ from which bound-state and scattering observables can be deduced~\cite{Atkinson:2020} (see App.~\ref{app:dom} for more details). 
The energy dependence of the self-energy ensures that many-body correlations manifest in $G_{\ell j}(r,r';E)$, providing a description beyond that of a mean field.
These correlations can be understood through the hole spectral function, defined as
\begin{equation}
   S_{\ell j}^h(r,r';E)=\frac{1}{\pi}\textrm{Im}G_{\ell j}^h(r,r';E).
\label{eq:spectral_function}
\end{equation}
The spectral function reveals that the strength of a given $\ell j$ shell can be fragmented over a wide range of energies, contrary to the mean-field picture of fully-occupied shells located at their respective mean-field energy levels (see Refs.~\cite{Atkinson20,Atkinson:2018,Atkinson:2019,Atkinson:2020} for explicit examples).
Results from DOM fits of $^{12}$C, $^{40}$Ca, $^{48}$Ca, and $^{208}$Pb are considered here. 


Traditionally, DOM fits are constrained  
by quasihole energies, particle numbers, charge densities, and, because of the dispersion relation, by all relevant scattering data up to 200 MeV. Here, we extend the treatment to incorporate also the total binding energy of each nucleus as obtained from the Green's function. 
A position-dependent energy density within the nucleus can then be defined such that its volume integral is the total binding energy. This approach provides a novel determination of nuclear energy densities based entirely on experimental data. Unlike mean-field or DFT energy densities, this approach is not constrained by prescribed analytics on energy densities. 
DOM fits produce occupation numbers that are not step-like, hence the corresponding kinetic-energy densities are not of a free-Fermi gas nature.
Moreover, these energy densities can be used to relate the energy of these nuclei to SCGF calculations in NM that only treat the consequences of SRC while including full off-shell propagation~\cite{Baldo:2012,Rios:2014}. 


The binding energy of a nucleus can be expressed as the expectation value of the Hamiltonian using the full $A$-body wave function, 
   $E_0^A = \braket{\Psi_0^A|\hat{H}|\Psi_0^A}$.
The energy density, $\mathcal{E}_A(r)$, of a nucleus can then be defined such that
\begin{equation}
  E_0^A = \int d^3r \mathcal{E}_A(r) = 4\pi\int_0^{\infty}drr^2\mathcal{E}_A(r).
   \label{eq:energy_density}
\end{equation}
The energy of the ground state can be recast into
the Migdal-Galitski sum rule~\cite{Galitski:1958} for both proton and neutron contributions with $E_0^A = E_0^N + E_0^Z$~\cite{Exposed!}.  
Since the DOM is calculated in a coordinate-space basis of Lagrange
functions~\cite{Baye_review}, $\mathcal{E}_A(r)$ can be calculated using 
\begin{align}
   \mathcal{E}_A(r) = \frac{1}{2}\int_0^{\varepsilon_F}&\sum_{\ell j}(2j+1)\left[\vphantom{\int}ES_{\ell j}^h(r,r;E)\right. \nonumber \\
   + \int_0^\infty & \left.\!\!\!dr'\ r'^2  \braket{r|\hat{T}_\ell|r'}  S_{\ell j}^h(r',r;E)\vphantom{\int}\right]dE,
  \label{eq:energy_density_sum}
\end{align}
where $\hat{T}_\ell$ is the kinetic-energy operator in the partial-wave basis. The first term corresponds to a
combination of the kinetic- and potential-energy densities~\cite{Exposed!} while
the second term represents the kinetic-energy density
\begin{equation}
   \mathcal{T}(r) =
   \sum_{\ell j}(2j+1)\mathcal{T}_{\ell j}(r),
   \label{eq:kinetic_density} 
\end{equation} 
where
\begin{equation}
   \mathcal{T}_{\ell j}(r) =
   \int_0^{\varepsilon_F}dE\int_0^\infty dr'r'^2\braket{r|\hat{T_\ell}|r'}S_{\ell j}^h(r',r;E).
   \label{eq:kinetic_density_lj} 
\end{equation} 
The volume integral of $\mathcal{T}(r)$ is the total kinetic energy
of the nucleus.  The kinetic-energy operator in coordinate space, 
\begin{equation} \braket{\bm{r}|\hat{T}|\bm{r}'} =
   \delta^3(\bm{r}-\bm{r}')\frac{-\hbar^2\bm{\nabla}_r^2}{2\mu}
   \label{eq:kinetic-operator} 
\end{equation}   
is used to calculate $\mathcal{T}(r)$, resulting in the following expression:
\begin{equation}
   r^2\mathcal{T}_{\ell j}(r) =
   \frac{-\hbar^2}{2\mu}\left[\frac{d^2}{dr^2} - \frac{\ell(\ell+1)}{r^2}\right]\left[rn_{\ell j}(r,r')r'\right]\Bigr\rvert_{r'=r},
   \label{eq:kinetic_explicit} 
\end{equation} 
where $n_{\ell j}(r,r')$ is the one-body density matrix defined as
\begin{equation}
   n_{\ell j}(r,r') = \int_0^{\varepsilon_F}dES_{\ell j}^h(r,r';E).
\label{eq:density}
\end{equation}

It is important to note that this derivation assumes there are no three-body terms in the nuclear interaction~\cite{Carbone:2013}.
The presence and need of a nuclear three-body force is undisputed~\cite{Carlson:2015}, but the arguments below do not change in any essential way by the assumption that Eq.~\eqref{eq:energy_density_sum} can be treated as exact (see Sec.~\ref{sec:abin} for further discussion). 
In particular, we will show that Variational Monte Carlo (VMC) calculations leading to exact Green's function Monte Carlo results (GFMC)~\cite{Pieper:2001} require only a modest attractive three-body contribution to the binding energy of light nuclei. 
With chiral interactions~\cite{Machleidt:2011}, the three-body force is important to generate NM saturation, but the many different versions hamper uniform conclusions and their softness may yield interior densities that are too large~\cite{Hagen:2012}. 

With Eq.~\eqref{eq:energy_density}, the binding energy of nuclei are also included in DOM fits with an accuracy of about 1.5\% and shown for $^{12}$C, $^{40}$Ca, $^{48}$Ca, and $^{208}$Pb in Table~\ref{table:sumrules}. Details of the $^{12}$C DOM fit are presented in the Appendix while details for $^{40}$Ca, $^{48}$Ca, and $^{208}$Pb fits can be found in Refs.~\cite{Atkinson:2018,Atkinson:2019,Atkinson:2020}, respectively.
\begin{table}[bt]
   \caption{Comparison of the DOM calculated binding
      energies of $^{12}$C, $^{40}$Ca, $^{48}$Ca, and $^{208}$Pb calculated using Eq.~\eqref{eq:energy_density_sum} to those
      calculated using the empirical mass formula. We use the parameters  $a_V=15.6$, $a_S=17.2$, $a_C=0.697$, and $a_A=46.6$ (all in MeV) in Eq.~\eqref{eq:mass_formula}. The experimental binding energies are
   shown in the last column. All listed energies are in MeV.}
   \label{table:sumrules}
         \begin{tabular}{cccc} 
            \hline
            \hline
            $A$ & DOM $E_0^A/A$\hspace{.2cm} & Mass Eq.\hspace{.2cm} & Exp. $E_0^A/A$ \\
            \hline
            \\[-10pt]
            $^{12}$C & -7.85 & -7.29 & -7.68\\
            \hline
            \\[-10pt]
            $^{40}$Ca & -8.46 & -8.50 & -8.55\\
            \hline
            \\[-10pt]
            $^{48}$Ca & -8.66 & -8.59 & -8.66 \\
            \hline
            \\[-10pt]
            $^{208}$Pb & -7.76 & -7.81 & -7.87 \\
            \hline
            \hline
         \end{tabular}
\end{table}
The agreement with experiment in Table~\ref{table:sumrules} is of a similar quality to that obtained by an empirical mass-formula fit. However, the DOM also reproduces the experimental charge densities, indicating that the hole spectral functions are well constrained. 

The energy density of $^{40}$Ca, weighted by the volume element $4\pi r^2$, and its separation in kinetic- and potential-energy density are shown in Fig.~\ref{fig:edens_40}. 
The weighting is chosen to emphasize the parts of
the energy density that contribute to the integral in Eq.~\eqref{eq:energy_density}. 
The figure clearly illustrates that the interior of the nucleus has a suppressed importance for the total energy on account of the phase space factor.
The nucleon point-density is shown in addition to the energy densities in Fig.~\ref{fig:edens_40} to demonstrate that the radial dependence of the energy density, $\mathcal{E}_A(r)$, and of the actual matter density, $\rho(r)$, are very similar. We explore this point further in the following section.

\section{Comparison with \textit{ab initio} calculations}
\label{sec:abin}
\begin{figure}[tb]
         \includegraphics[scale=1.0]{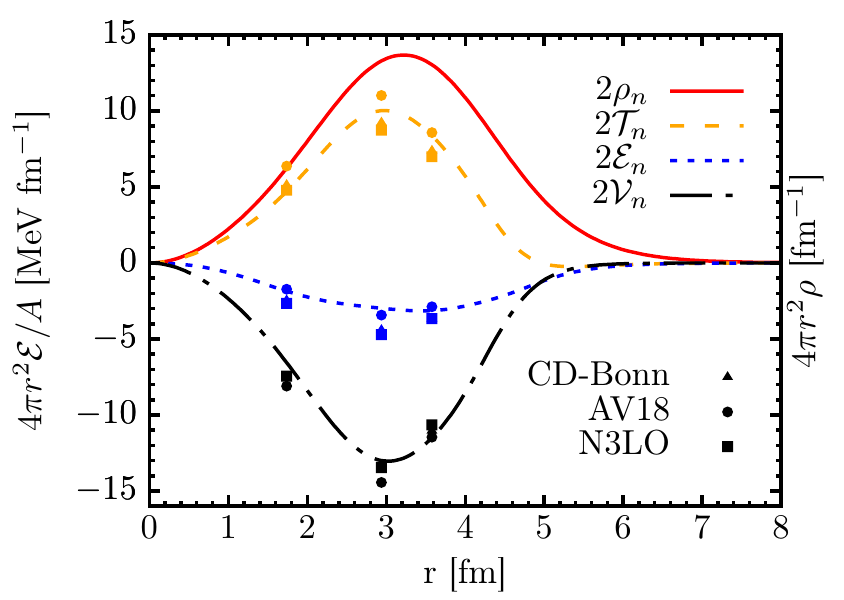}
   \caption[The energy density of $^{40}$Ca]{
      Energy densities in $^{40}$Ca calculated from the DOM using Eq.~\eqref{eq:energy_density_sum}. Each line corresponds to twice the contribution from neutrons (see text). The curves correspond to the binding-energy density (dotted line), kinetic-energy density (dashed line), potential-energy density (dot-dashed line), and nucleon point-density (solid line). All curves
      are weighted by a volume element $4\pi r^2$. The points are taken from a SCGF calculation in NM for three different interactions based on Ref.~\cite{Rios:2014} at densities corresponding to 0.08, 0.12, and 0.16 fm$^{-3}$. 
   }
   \label{fig:edens_40}
\end{figure}

SCGF calculations in NM from Ref.~\cite{Rios:2014} are represented by symbols in
Fig.~\ref{fig:edens_40}. Each different symbol corresponds to a different NN
interaction in the SCGF calculation, where the triangles correspond to the charge-dependent Bonn (CD-Bonn)
interaction~\cite{Machleidt:1996}, the circles correspond to the Argonne $v_{18}$ (AV18)
interaction~\cite{AV18:1995}, and the squares correspond to the Idaho next-to-next-to-next-to-leading order (N3LO) chiral interaction~\cite{Entem:2003}. The calculation in NM is for specific
values of the nuclear density which are mapped to radii using the DOM matter density.
These results cannot be directly compared to the energy density in finite nuclei because there is no Coulomb force included in NM.
Since there are an equal number of protons and neutrons in
$^{40}$Ca, isospin symmetry implies that their distributions would be the same if the Coulomb force were ignored. Thus, using twice the neutron energy density in $^{40}$Ca is an effective way
of removing the influence of the Coulomb force. This is how the lines in Fig.~\ref{fig:edens_40} are generated. 
These isospin-corrected results provide energy densities that are similar to those predicted by SCGF calculations with very different NN interactions. The agreement with the NM
calculations is striking since the latter only include effects of short-range (SRC) and tensor correlations as suggested in Ref.~\cite{Dewulf:2003}. This implies that the interior of $^{40}$Ca exhibits NM-like properties. 

The interaction with the best agreement with the DOM energy density in Fig.~\ref{fig:edens_40} is AV18. It is
interesting that, unlike the other two interactions, the harder AV18 correctly reproduces the nuclear saturation density
$\rho_0\approx 0.16$~fm$^{-3}$~\cite{Bethe:1971,Myers:1996} in the SCGF calculation reported in Ref.~\cite{Baldo:2012}, but saturates at about $-11.5$~MeV. 
This is in disagreement with the canonical value, $a_V \approx -16$ MeV, which comes from the empirical mass formula.
However, it is clear from Fig.~\ref{fig:edens_40} that the interior of the
nucleus does not determine the binding energy since it minimally contributes to Eq.~\eqref{eq:mass_formula}.
Concurrently, it is apparent that SCGF calculations in NM generate quantitatively correct binding-energy densities as shown in Fig.~\ref{fig:edens_40}.
As discussed above, the interior of the nucleus saturates around $\rho_0$,
implying that this region corresponds to saturated NM. This is further
supported by the reproduction of the smooth interior charge density in the DOM.
We therefore draw three conclusions. First, with
the interpretation that NM is representative of the core of finite nuclei, 
we infer that there is no strong constraint that the binding energy of NM has to be $a_V$. Second, the agreement between the NM points and $^{40}$Ca in Fig.~\ref{fig:edens_40} is consistent with the fact that SRC are primarily what link finite nuclei to NM~\cite{Dewulf:2003,Dickhoff04,Dickhoff:2016}. 
Third, we conjecture that the AV18 interaction not only reproduces the saturation density, but produces a reasonable saturation energy ($E_0\approx -11.5$ MeV) given that the AV18 points in Fig.~\ref{fig:edens_40} are consistent with the DOM $^{40}$Ca energy density.  
This conjecture is empirically supported by the fact that
the AV18 + Urbana-IX~\cite{Pudliner:1995} (3-body interaction) was used to derive the
Akmal, Pandharipande, and Ravenhall (APR) EOS of
NM~\cite{APR:1998}. It is widely used in calculations of neutron star structure, all of which are consistent with current
observations of neutron stars including the recent neutron star merger event~\cite{LIGO:2017,APR:1998}. The APR EOS correctly predicts the value of $\rho_0$ but with a minimum energy of $E_0 = -12.6$ MeV.
While the value of this minimum energy has been seen as a defect of the APR EOS, its success in describing nuclear
systems further supports a saturation energy different from $a_V$.

\begin{figure}[tb]
         \includegraphics[scale=1]{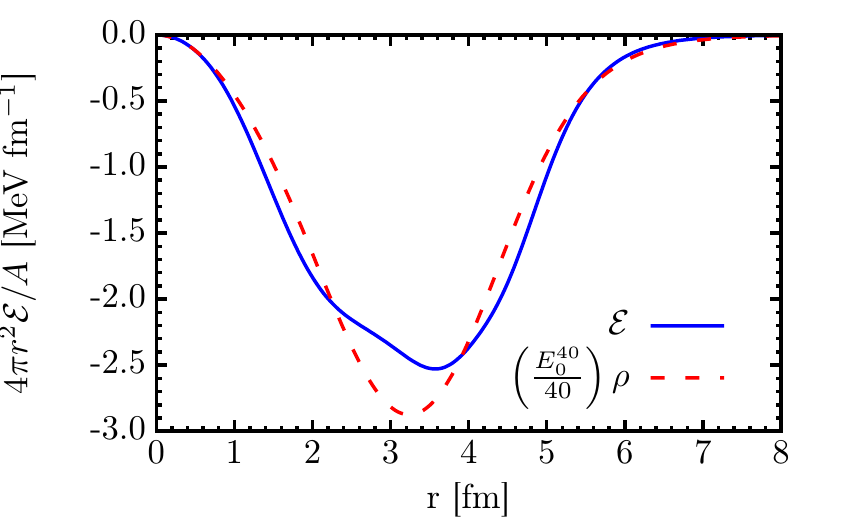}
   \caption[Comparison of the energy density to the scaled nucleon density in $^{40}$Ca.]{
   The binding-energy density of Eq.~\eqref{eq:energy_density_sum} (solid line) compared to the scaled nucleon density of Eq.~\eqref{eq:energy_relation} (dashed line) in $^{40}$Ca. 
   }
   \label{fig:ecomp}
\end{figure}

The fact that the binding energy density traces the matter density in Fig.~\ref{fig:edens_40} is not surprising when
considering the decomposition of the binding energy using full $A$-body wave functions,
\begin{align}
   E_0^A &= \braket{\Psi_0^A|\hat{H}|\Psi_0^A} = E_0^A \braket{\Psi_0^A|\Psi_0^A} \nonumber\\
    = E_0^A&\int d^3r_1 \left[\int d^3r_2...d^3r_A \left|\Psi_0^A(\bm{r}_1,\bm{r}_2,...,\bm{r}_A)\right|^2\right],
   \label{eq:energy_exact}
\end{align}
where the complete set $\{\ket{\bm{r}_1,\bm{r}_2,...,\bm{r}_A}\}$ has been inserted and all other quantum numbers are suppressed for clarity. Noting that the bracketed term in
Eq.~\eqref{eq:energy_exact} is the one-body density distribution $\rho(\bm{r})$, the binding energy can be written as
\begin{equation}
   E_0^A = \frac{E_0^A}{A} \!\! \int \!\! d^3r\rho_A(\bm{r}) 
 \!\!  \implies \!\! \mathcal{E}_A(r) = \left(\frac{E_0^A}{A}\right)\rho_A(r).
   \label{eq:energy_relation}
\end{equation}
Eq.~\eqref{eq:energy_relation} is not a unique expression of the energy density since only its
integral (the binding energy) is an observable.  
However, Eq.~\eqref{eq:energy_relation} is a
natural choice because the energy densities in Fig.~\ref{fig:edens_40} roughly trace the matter
density.  While Eq.~\eqref{eq:energy_relation} is exact, it cannot be used as a replacement for
Eq.~\eqref{eq:energy_density_sum} because there is no guarantee that the DOM propagator is equal to
the exact propagator, which would be built from the exact $A$-body ground-state wave
function~\cite{Exposed!}.  
This is demonstrated in
Fig.~\ref{fig:ecomp}, which shows the energy density in $^{40}$Ca calculated
using both Eq.~\eqref{eq:energy_density_sum} and Eq.~\eqref{eq:energy_relation}.
The general agreement of the curves in Fig.~\ref{fig:ecomp} is quantified by the
similarity of the rms radii of the displayed energy and scaled nucleon density
of 3.477 and 3.480 fm, respectively. This reveals that the DOM description of the
density is close to exact. It is not surprising that there are deviations, since
the DOM fit constrains the density which is only an indirect way of
constraining the full $A$-body wave function. 

\begin{figure}[b]
         \includegraphics[scale=1.0]{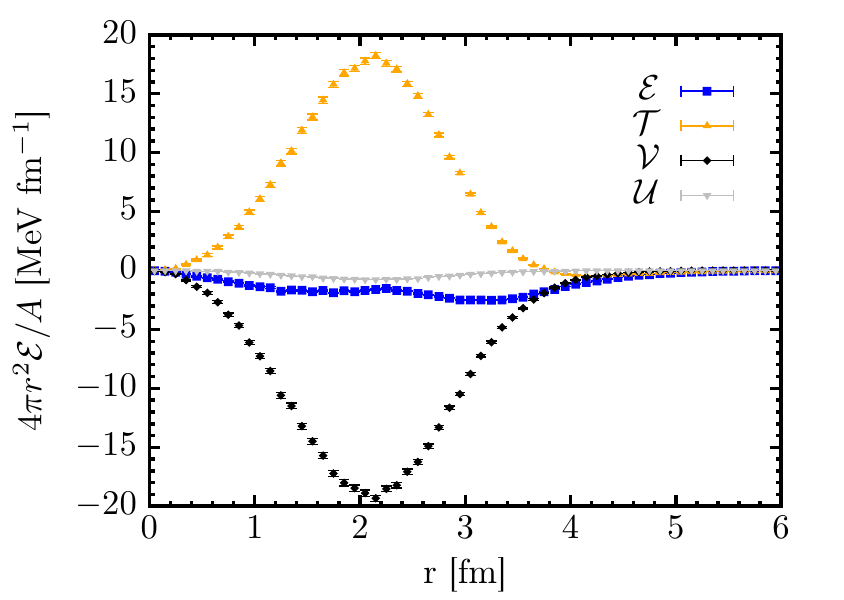}
   \caption{Results of a GFMC calculation of $^{8}$Be with $\mathcal{E}$, $\mathcal{T}$, $\mathcal{V}$, and $\mathcal{U}$ representing the total binding-energy density, the kinetic-energy density, the two-body potential-energy density, and the three-body potential-energy density, respectively.}
   \label{fig:gfmc}
\end{figure} 

\begin{figure}[h]
         \includegraphics[scale=1.0]{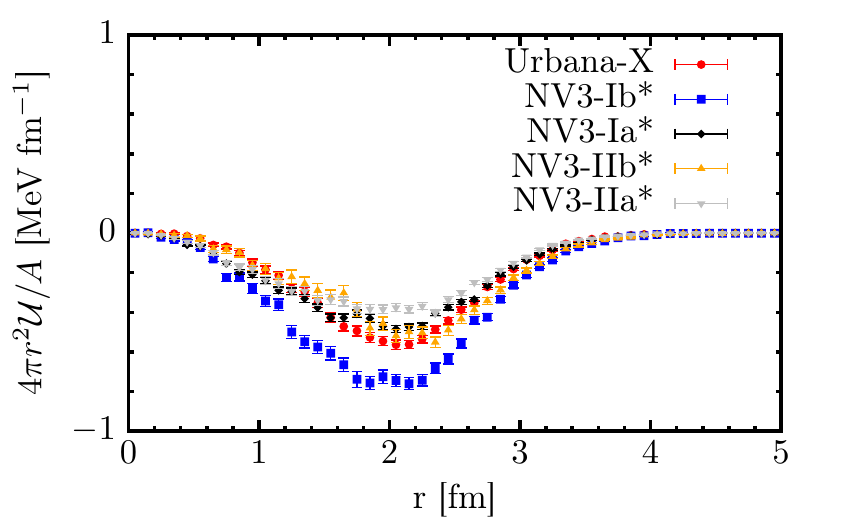}
   \caption[The 3-body energy density of $^{40}$Ca]{
    Illustration of the 3-body potential-energy densities for different chiral interactions~\cite{Piarulli:2016} and the UX~\cite{Wiringa:2014} for ${}^{12}$C. 
   }
   \label{fig:3-body}
\end{figure} 

A method that is well-suited to calculate the energy density using Eq.~\eqref{eq:energy_exact} is
GFMC. 
The results of a GFMC calculation for the $^{8}$Be binding-energy density is shown in Fig.~\ref{fig:gfmc}, generating a total kinetic energy of $239$ MeV, a two-body potential energy of $-287$ MeV, a three-body potential
energy of $-10.7$ MeV, and a total energy of $-56.1$ MeV compared to the experimental value of $-56.5$ MeV. In this calculation, the AV18 +
Urbana-X~\cite{Wiringa:2014} (UX) interactions were employed to generate the ground-state wave function. 
The results in Fig.~\ref{fig:gfmc} include the
contribution of the three-body interaction to the energy density. Comparing the two- and three-body potential density clarifies that the latter contributes modestly to the total energy
density and certainly is not capable of changing its shape. Consequently, we expect that ignoring the three-body interaction by using Eqs.~\eqref{eq:energy_density}  and \eqref{eq:energy_density_sum} in the DOM analysis will not alter the shape of the binding-energy density.

In order to further assess the effects of the NNN interaction, we also report VMC calculations of $^{12}$C
using the three-body components (NV3*) of the Norfolk chiral interactions (NV2+3*)~\cite{Piarulli:2016,Piarulli:2018,Baroni:2018} as well as the UX NNN interaction. In Fig.~\ref{fig:3-body}, we show the three-body
potential densities calculated in $^{12}$C using these five different interactions. In particular, NV3* models have been constrained by fitting the trinucleon energies and the empirical value of the Gamow-Teller matrix element in tritium $\beta$ decay in combination with the corresponding Norfolk two-body potential (NV2). There are two classes (I and II) of NV2, differing only in the range of laboratory energy over
which they are fitted to the nucleon-nucleon database; class I up to 125 MeV, and
class II up to 200 MeV. For each class, two combinations of short- and long-range regulators 
have been used, namely ($R_S$, $R_L$)=(0.8, 1.2) fm (models NV2-Ia and NV2-IIa) and 
($R_S$, $R_L$)=(0.7, 1.0) fm (models NV2-Ib and NV2-IIb). 
In Table.~\ref{table:potential-energy}, we explicitly report the potential energy contributions to the binding energy of $^{12}$C using the interactions displayed in Fig.~\ref{fig:3-body}.
Again, we find that the contributions from NNN forces (U) to the total energy (density) are small in comparison to the corresponding NN ones (V). 
\begin{table}[h]
   \caption{Potential energy contributions from the NNN interactions (U) and the corresponding NN interactions (V) shown in Fig.~\ref{fig:3-body} for $^{12}$C.}
   \label{table:potential-energy}
         \begin{tabular}{c c c} 
            \hline
            \hline
            Interaction & V & U \\
            \hline
            AV18+UX & -457 & -10.5 \\
            \hline
            NV2+3-Ib* & -383 & -15.9 \\
            \hline
            NV2+3-Ia* & -379 & -10.3 \\
            \hline
            NV2+3-IIb* & -416 & -10.4 \\
            \hline
            NV2+3-IIa* & -411 & -8.91 \\
            \hline
            \hline
         \end{tabular}
\end{table}
As expected, there is
some variation in the NNN potential densities for the different interactions used. However, the fact that
these variations are small demonstrates that, regardless of the NNN interaction used, the shape of
the binding-energy density is not altered in a significant way by NNN forces. 
In all cases, the NNN potential-energy-density contribution is small in comparison with the corresponding NN one. We expect that the conclusions drawn for $^{8}$Be and $^{12}$C in terms of relative sizes of V and U will hold for the heavier nuclei considered in the following section.



\section{Analysis}
\label{sec:analysis}
The nuclear energy density can be further explored for the heavier nuclei $^{48}$Ca and $^{208}$Pb. The agreement between
Eq.~\eqref{eq:energy_relation} and Eq.~\eqref{eq:energy_density_sum} in $^{48}$Ca and $^{208}$Pb is comparable to that of $^{40}$Ca.
The case of $^{208}$Pb
is particularly interesting because the interior is more extended than in $^{40}$Ca and $^{48}$Ca.  This implies that
finite-size (surface) effects are reduced in this region of $^{208}$Pb, making it an even more suitable system to compare with NM.
Using isospin symmetry to remove the effect of the Coulomb interaction on the energy density of $^{40}$Ca
is not valid in $^{208}$Pb, since $N > Z$.  
While removing the Coulomb energy density from $\mathcal{E}(r)$ would provide a
NM-like energy density, the Coulomb potential is still
reflected in the matter density of $^{208}$Pb (see also Ref.~\cite{Atkinson:2020}).
One way to compare with the NM
calculations for asymmetric matter from Ref.~\cite{Rios:2014} is to completely remove the Coulomb potential from the DOM
self-energy. 
To preserve the proton number, the proton Fermi energy must therefore be shifted such that it remains between the particle-hole gap of the protons.
The resulting Coulomb-less matter density exactly confirms the expected 0.16~fm$^{-3}$ in the interior of $^{208}$Pb.

\begin{figure}[tb]
         \includegraphics[scale=1]{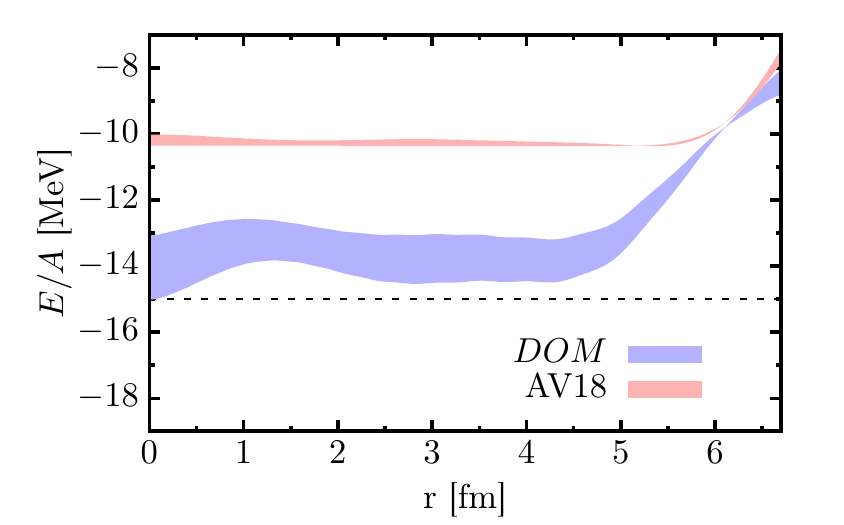}
   \caption[Binding energy as a function of radius in $^{208}$Pb.]{
      Binding energy as a function of radius in $^{208}$Pb. 
      The thick blue band covers the range of energies of $^{208}$Pb calculated using the
      DOM matter density (top) and the use of the DOM proton density
      scaled by 208/82 (bottom), both with Coulomb removed. 
      The narrow band is similarly obtained from the SCGF calculations for the AV18~\cite{Rios:2014} (see text). 
      The dashed line is the expected energy from the empirical mass formula.
   }
   \label{fig:energy_208}
\end{figure} 

The energy in the interior can be approximately calculated from the energy density using
Eq.~\eqref{eq:energy_relation}, 
\begin{equation} 
   E_A(r) \approx \mathcal{E}_A(r)\left(\frac{A}{\rho_A(r)}\right).
   \label{eq:energy_approx} 
\end{equation}
This approximation should be valid for small values of $r$, where the nuclear density is
relatively constant and saturated. 
The binding energy with Coulomb removed as a function of $r$ in $^{208}$Pb is shown in Fig.~\ref{fig:energy_208}.  
The ambiguity to determine the Coulomb-less interior density is reflected in the wide band.
The thin band represents the interpolation of SCGF calculations from Ref.~\cite{Rios:2014} using AV18 at densities corresponding to 0.08, 0.12, and 0.16 fm$^{-3}$ obtained in the same way. 
These NM results require an additional 2-3 MeV per particle attraction to reproduce the DOM result, which is not inconsistent with the trend obtained for the required contribution of the three-body interaction to accurately describe the energies of light nuclei with many-body methods~\cite{Carlson:2015,Hoppe:2019}.
Additional binding might result from LRC in heavier nuclei which are not accounted for by the SRC results depicted in Fig.~\ref{fig:energy_208} for AV18.
The contribution of the symmetry energy per nucleon from the empirical mass formula in $^{208}$Pb is $E_{sym} = 1.04$ MeV, leading to the expectation of the interior energy
of $^{208}$Pb to be $E_0^{208} = -15.0$ MeV based on the empirical mass formula (see dashed line in Fig.~\ref{fig:energy_208}).
Our analysis therefore suggests that the
energy in the interior 
(and hence the saturation energy)
is less bound than what is expected from the empirical mass formula. In $^{208}$Pb, we find $E_A/A \approx -14$ MeV. 


\begin{figure}[bth]
         \includegraphics[scale=1]{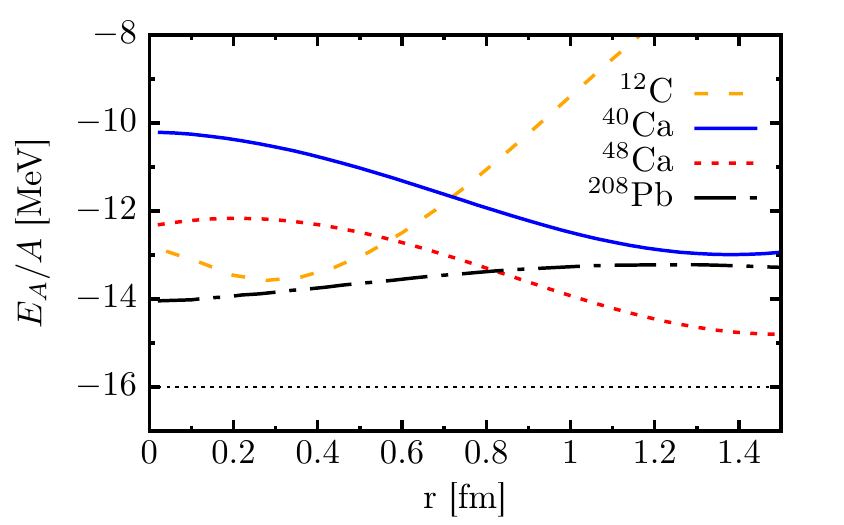}
   \caption[Binding energy as a function of radius in $^{40}$Ca, $^{48}$Ca, and $^{208}$Pb.]{
      Binding energy as a function of radius in $^{12}$C (dashed line), $^{40}$Ca (solid line), $^{48}$Ca (dotted line), and $^{208}$Pb (dot-dashed line).
      The latter reflects the middle of the band in Fig.~\ref{fig:energy_208}. The canonical -16 MeV$/A$ binding is also shown.
   }
   \label{fig:energy_all}
\end{figure}


A comparison of the DOM energy as a function of radius for $^{12}$C, $^{40}$Ca, $^{48}$Ca, and $^{208}$Pb is shown in Fig.~\ref{fig:energy_all}, where the
Coulomb contribution has been removed from each nucleus. The energies in the core of each nucleus are all within a few MeV of each other. Near the origin, all of them are significantly less bound than $16$~MeV per particle. We expect that this result holds across a wide range of isotopes. It also appears to be robust to statistical uncertainties in the DOM fits and, as discussed above, to the (relatively small) contribution of NNN forces. We take this as a strong hint that the central energy density departs significantly from the canonical value obtained via the $a_V$ parameter of mass formulae.

\section{Conclusions and outlook}
\label{sec:conclusions}

The interpretation that the interior of the
nucleus is a close approximation to NM implies that a macroscopic mass formula, such as
Eq.~\eqref{eq:mass_formula}, is not a suitable way of determining the binding energy of NM at
saturation. Our results invalidate this approach by shedding light on two different aspects.
First, Fig.~\ref{fig:edens_40} clearly shows that the interior of the nucleus does not significantly contribute to the total binding energy. Nuclear masses should thus only have small contributions from the saturated, deep nuclear interior. In other words, mass formulae are unlikely to capture the energy dynamics of the nuclear interior, including its mass number dependence.
Second, the interior saturation energies, as shown by the DOM analysis above, do not necessarily agree with the value of $a_V$ that provides a good fit to nuclear masses. 
It has been noted in the past~\cite{Dewulf:2003,Dickhoff04,Dickhoff:2016} that LRC in finite nuclei
and NM are not commensurate, implying an uncertainty in the extrapolation from
Eq.~\eqref{eq:mass_formula} to NM. 
Taking our results into consideration leads to the inevitable conclusion that the saturation energy
of symmetric NM is less than the canonical value of $16$~MeV per particle. Considering the interior of
$^{208}$Pb indicates that $E_0$ is actually closer to 13-14 MeV. This is also closer to the value generated by SCGF calculations of NM with the AV18 force.

These results can also be interpreted in terms of different energy values that are traditionally expected to be similar. On the one hand, $a_V$ quantifies the bulk mass-number dependence of nuclear binding energies. On the other, the saturation energy of NM, $E_0$, provides the minimum energy of an infinite system. Mass-number fits are however performed on finite nuclei data and thus extrapolations to the $A \to \infty$ limit need to be considered with care~\cite{Dobaczewski:2002}. 
Experience with other many-body systems like Helium drops indicates that one may be able to shift contributions of different $A-$dependent terms within mass formulae, thus changing the value of $a_V$. Our analysis in fact suggests that the value of $E_0$ may be about $10 \%$ smaller than that obtained from $a_V$. It remains to be seen whether mass formulae with lower values of $a_V$ provide quantitative fits to nuclear masses.  

To our knowledge, the systematic  uncertainty in the value of $E_0$ extrapolated from $a_V$ has not been investigated since the construction of Eq.~\eqref{eq:mass_formula}.
With the development of more precise NN+NNN interactions as well as the continued improvement of many-body methods, it is important to have an accurate value of the nuclear saturation point. This is often used in benchmarking NN and NNN forces \cite{Drischler:2019}. In fact, modern chiral
interactions already incorporate nuclear observables, such as binding energies and charge radii of nuclei, to their fitting protocols~\cite{Ekstrom:2015}. It has been suggested that the NM saturation point should also be added to these fits~\cite{Hagen:2014,Jiang:2020,Drischler:2019}. In light of this and the conclusions of this article, it is imperative that new methods of determining the value and uncertainty of $E_0$ are explored. 

We suggest a way forward in connecting $E_0$ to nuclear observables. Rather than relying uniquely on bulk masses, we use the  energy density in the nuclear interior, $E_A(\rho)$, to provide an estimate for $E_0$. The energy density is accessible by several contemporary many-body methods. Here, for instance, we have used quantum Monte Carlo simulations in light nuclei to validate the DOM predictions and gauge the importance of different components to the energy density. This has helped confirm that the contribution of NNN is relatively small. A similar analysis could be performed with other \textit{ab initio} methods that can reach higher masses and even compute NM within the same footing~\cite{Hagen:2014}. This would provide a theory-to-theory connection between the saturation point of NM and the properties of nuclei.

Our results also suggest that in addition to purely theoretical methods, nuclear data can also provide an insight into the energy density profile within nuclei. The unified view of nuclear scattering data and bound properties obtained from the DOM is in fact able to provide a quantitative description of the nuclear energy density. In this first exploratory work, we have not dealt explicitly with NNN forces, but some steps in this direction could be easily explored in conjunction with similar many-body methods like the SCGF approach. Extending the DOM fits to other isotopes across the nuclear chart (already begun in Refs.~\cite{Pruitt:2020,Pruitt:2020C}) will also provide a further quantitative, nuclear-data-inspired understanding of the mass evolution of nuclear energy densities.

\section*{Acknowledgments}
The work of MCA and WHD was supported by the U.S. National Science Foundation
under grants PHY-1613362 and PHY-1912643, the work of AR by the UK Science and
Technology Facilities Council (STFC) through grant ST/P005314/1, the work of
RBW by the U.S. Department of Energy, Office of Nuclear Physics under contract
DE-AC02-06CH11357 and the NUCLEI SciDAC program, with computational resources
provided by the Argonne Laboratory Computing Resource Center, and the work of
MP by the U.S. Department of Energy funds through the FRIB Theory Alliance
award DE-SC0013617 with computational resources provided by the Argonne
Leadership Computing Facility via the 2019/2020 ALCC ``Low energy
neutrino-nucleus interactions" for the project NNInteractions. TRIUMF receives federal funding via a contribution agreement with the National Research Council of Canada.

\appendix

\section{The DOM}
\label{app:dom}

 It was recognized long ago that the irreducible self-energy represents the potential 
 that describes elastic-scattering observables~\cite{Bell59}. 
 The link with the potential at negative energy is then provided by the Green's function framework as was realized by Mahaux and Sartor who introduced the DOM as reviewed in Ref.~\cite{Mahaux91}. 
 The analytic structure of the nucleon self-energy allows one 
 to apply the dispersion relation, which relates the real part of the self-energy at a given energy to a dispersion integral of its imaginary part over all energies.
 The energy-independent correlated Hartree-Fock (HF) contribution~\cite{Exposed!} is removed by employing a subtracted dispersion relation with the Fermi energy ($\varepsilon_F$) used as the subtraction point~\cite{Mahaux91}.
 The subtracted form has the further advantage that the emphasis is placed on energies closer to the Fermi energy for which more experimental data are available.
 The real part of the self-energy at the Fermi energy is then still referred to as the HF term, but is sufficiently attractive to bind the relevant levels.
 In practice, the imaginary part is assumed to extend to the Fermi energy on both sides while being very small in its  vicinity.
 The subtracted form of the dispersion relation employed in this work is given by
 \begin{align}
    \textrm{Re}\ \Sigma^*(\alpha,\beta;E) &= \textrm{Re}\
    \Sigma^*(\alpha,\beta;\varepsilon_F) \label{eq:dispersion} \\ -
    \mathcal{P}\int_{\varepsilon_F}^{\infty} \!\! \frac{dE'}{\pi}&\textrm{Im}\
    \Sigma^*(\alpha,\beta;E')[\frac{1}{E-E'}-\frac{1}{\varepsilon_F-E'}] \nonumber
    \\ + \mathcal{P} \! \int_{-\infty}^{\varepsilon_F} \!\!
    \frac{dE'}{\pi}&\textrm{Im}\
    \Sigma^*(\alpha,\beta;E')[\frac{1}{E-E'}-\frac{1}{\varepsilon_F-E'}],
    \nonumber      
 \end{align}
 where $\mathcal{P}$ is the principal value. 
 The static term is denoted by  $\Sigma_{\text{HF}}$ from here on. 
 Equation~\eqref{eq:dispersion} constrains the real part of the self-energy through empirical information of the HF term and empirical knowledge of the imaginary part, which is closely tied to experimental data. 
 Initially, standard functional forms for these terms were introduced by Mahaux and Sartor who also cast the DOM potential in a local form by a standard transformation which turns a nonlocal static HF potential into an energy-dependent local potential~\cite{Perey:1962}.
 Such an analysis was extended in Refs.~\cite{Charity06,Charity:2007} to a sequence of Ca isotopes and in Ref.~\cite{Mueller:2011} to semi-closed-shell nuclei heavier than Ca.
 The transformation to the exclusive use of local potentials precludes a proper calculation of nucleon particle number and expectation values of the one-body operators, like the charge density in the ground state. 
 This obstacle was eliminated in Ref.~\cite{Dickhoff:2010}, but it was shown that the introduction of nonlocality in the imaginary part was still necessary in order to accurately account for particle number and the charge density~\cite{Mahzoon:2014}.
 Theoretical work provided further support for this introduction of a nonlocal representation of the imaginary part of the self-energy~\cite{Waldecker:2011,Dussan:2011}.
 A recent review has been published in Ref.~\cite{Dickhoff:2017}.

 We implement a nonlocal representation of the self-energy following
 Ref.~\cite{Mahzoon:2014} where $\Sigma_{\text{HF}}(\bm{r},\bm{r'})$ and
 $\textrm{Im}\ \Sigma(\bm{r},\bm{r'};E)$ are parametrized, using Eq.~\eqref{eq:dispersion} to generate the energy dependence of the
 real part. The HF term consists of a volume term, spin-orbit term,  and a wine-bottle-shaped term~\cite{Brida11} to simulate a surface contribution. The imaginary self-energy consists of volume, surface, and spin-orbit terms (see App.~\ref{app:param}). 
Nonlocality is represented using the Gaussian form 
 \begin{equation}
    H(\bm{s},\beta) = \pi^{-3/2}\beta^{-3}e^{-\bm{s}^2/\beta^2} ,
    \label{eq:nonlocality}
 \end{equation}
 where $\bm{s} = \bm{r} -\bm{r}'$, 
 as proposed in Ref.~\cite{Perey:1962}. 
 As mentioned previously, it was customary in the past to replace nonlocal potentials by local, energy-dependent potentials~\cite{Mahaux91,Perey:1962,Fiedeldey:1966,Exposed!}. The introduction of an energy dependence alters the dispersive
 correction from Eq.~\eqref{eq:dispersion} and distorts the normalization, leading to incorrect spectral functions and related quantities~\cite{Dickhoff:2010}. Thus, a nonlocal implementation permits the self-energy to accurately 
 reproduce important observables such as the charge density and particle number. 

 The potential is transformed from coordinate space to a Lagrange basis using Legendre and Laguerre polynomials for scattering and bound states, respectively~\cite{Baye_review}.
  The propagator is found by inverting the Dyson equation,
 \begin{align}
    G_{\ell j}(\alpha,\beta;E) &= G_{\ell}^{(0)}(\alpha,\beta;E) \nonumber \\ +
    \sum_{\gamma,\delta}&G_{\ell}^{(0)}(\alpha,\gamma;E)\Sigma_{\ell
    j}^*(\gamma,\delta;E)G_{\ell j}(\delta,\beta;E) ,
    \label{eq:dyson}
 \end{align}
 while all scattering calculations are done in the framework of $R$-matrix theory~\cite{Baye:2010}. 
 Implementations of the nonlocal DOM in $^{40}$Ca, $^{48}$Ca, and $^{208}$Pb have previously been published in Refs.~\cite{Mahzoon:2017,Atkinson:2018,Mahzoon:2014,Atkinson:2020}.

\section{Parametrization of the potentials}
\label{app:param}

We provide a detailed description of the parametrization of the proton and neutron self-energies in
$^{12}$C used in the fits to bound and scattering data. The parametrizations of $^{40}$Ca, $^{48}$Ca, and $^{208}$Pb can be found in Refs.~\cite{Atkinson:2018,Atkinson:2019,Atkinson:2020}, respectively. The $\pm$ in superscripts and subscripts refer to above ($+$) and below ($-$) the Fermi
energy, $\varepsilon_F$. 

We restrict the nonlocal contributions to the HF term and to the volume and surface contributions to the imaginary part of the potential.
We write the HF self-energy term in the following form with the local Coulomb contribution.
\begin{eqnarray}
\Sigma_{HF}(\bm{r},\bm{r}') = \Sigma^{nl}_{HF}(\bm{r},\bm{r}') + V^{nl}_{so}(\bm{r},\bm{r}') + \delta(\bm{r}-\bm{r}') V_C(r) , \nonumber \\
\nonumber
\end{eqnarray}

The nonlocal term is split into a volume and a narrower Gaussian term of opposite sign to make the final potential have a wine-bottle shape.

\begin{eqnarray}
\Sigma_{HF}^{nl}\left( \bm{r},\bm{r}' \right) = -V_{HF}^{vol}\left( \bm{r},\bm{r}'\right) 
+ V_{HF}^{wb}(\bm{r},\bm{r}') ,
\label{eq:HFn}
\nonumber
\end{eqnarray}
where the volume term is given by
\begin{equation}
\begin{split}
V_{HF}^{vol}\left( \bm{r},\bm{r}' \right) =  V^{HF}
\,f \left ( \tilde{r},r^{HF}_{},a^{HF} \right ) \\ \times
 \left [ x H \left( \bm{s};\beta^{vol_1} \right) + (1-x) H \left( \bm{s};\beta^{vol_2}\right) \right ] \\
\end{split} \label{eq:HFvol} 
\end{equation}
allowing for two different nonlocalities with different weights ($0 \le x \le1$). 
With   the notation $\tilde{r} =(r+r')/2$ and $\bm{s}=\bm{r}-\bm{r}'$,
the wine-bottle ($wb$) shape is described by
\begin{equation}
V_{HF}^{wb}(\bm{r},\bm{r}') = V^{wb}_{}  \exp{\left(- \tilde{r}^2/(\rho^{wb})^2\right)} H \left( \bm{s};\beta^{wb} \right ),
\label{eq:wb}
\end{equation}
where $H(\bm{s},\beta)$ is given in Eq.~\eqref{eq:nonlocality}.
As usual, we employ a Woods-Saxon shape
\begin{eqnarray}
f(r,r_{i},a_{i})=\left[1+\exp \left({\frac{r-r_{i}A^{1/3}}{a_{i}}%
}\right)\right]^{-1} .
\label{Eq:WS}
\end{eqnarray}
The Coulomb term is obtained from the charge density distribution in the standard way~\cite{Jackson}.

The spin-orbit potential has the following form, 
\begin{equation}\begin{split}
      V^{nl}_{so}(\bm{r},\bm{r'})= \left( \frac{\hbar}{m_{\pi }c}\right)
^{2} V^{so}\frac{1}{\tilde{r}}\frac{d}{d\tilde{r}}f(\tilde{r},r^{so},a^{so})\; \bm{\ell}\cdot \bm{\sigma} \\
\times H(\bm{s};\beta^{so}), \end{split}
\label{eq:HFso}
\end{equation}
where $\left( \hbar /m_{\pi }c\right) ^{2}$=2.0~fm$^{2}$ 
as in Ref.~\cite{Mueller:2011}.

The introduction of nonlocality in the imaginary part of the self-energy is well-founded theoretically both for long-range correlations~\cite{Waldecker:2011} as well as in short-range ones~\cite{Dussan:2011}. 
Its implied $\ell$-dependence is essential in reproducing the correct particle number for protons and neutrons.
The fully-nonlocal imaginary part of the DOM self-energy has the following form,
\begin{align}
\label{eq:imnl}
&\textrm{Im}\ \Sigma^{nl}(\bm{r},\bm{r}';E) = \hspace{5cm} \\ \nonumber  
&-W^{vol}_{0\pm}(E) f\left(\tilde{r};r^{vol}_{\pm};a^{vol}_{\pm}\right)H \left( \bm{s}; \beta_{\pm}^{vol}\right) \hspace{1cm} \\ \nonumber
& + 4 a^{sur}_{sym} W^{sur}_{\pm}(E) H \left( \bm{s};\beta^{sur}_{\pm} \right ) \frac{d}{d \tilde{r} }f(\tilde{r},r^{sur}_{\pm},a^{sur}_{\pm}) \\ \nonumber 
&+ \textrm{Im}\Sigma_{so}(\bm{r},\bm{r}';E).
\end{align}
Note that the parameters relating to the shape of the imaginary spin-orbit term
are the same as those used for the real spin-orbit term.
At energies well removed
from $\varepsilon_F$, the form of the imaginary volume potential should not be
symmetric about $\varepsilon_F$ as indicated by the $\pm$ notation in the subscripts and superscripts~\cite{Dussan:2011}.
While more symmetric about $\varepsilon_F$, we have allowed a similar option for the surface absorption that is also supported by theoretical work reported in Ref.~\cite{Waldecker:2011}.

Allowing for the aforementioned asymmetry around $\varepsilon_F$ the following form was assumed for 
the depth of the volume potential~\cite{Mueller:2011}
\begin{widetext} 
\begin{equation}
W^{vol}_{0\pm}(E) =  \Delta W^{\pm}_{NM}(E) +  
\begin{cases}
0 & \text{if } |E-\varepsilon_F| < \mathcal{E}^{vol} \\
\left [ A^{vol} \pm \eta^{vol} \right ]  \frac{\left(|E-\varepsilon_F|-\mathcal{E}^{vol}\right)^4}
{\left(|E-\varepsilon_F|-\mathcal{E}^{vol}\right)^4 + (B^{vol})^4} & 
 \text{if } |E-\varepsilon_F| > \mathcal{E}^{vol} ,
\end{cases} 
\label{eq:volumeS}
\end{equation}
\end{widetext}
where $\Delta W^{\pm}_{NM}(E)$ is the energy-asymmetric correction modeled after
nuclear-matter calculations. The asymmetry above and below $\varepsilon_F$ is essential to accommodate the Jefferson Lab $(e,e'p)$ data at large missing energy.
The energy-asymmetric correction was taken as 
\begin{widetext} 
\begin{equation}
\Delta W^{\pm}_{NM}(E)=
\begin{cases}
\alpha \left [A^{vol}_{+} \pm \eta^{vol} \right ]\left[ \sqrt{E}+\frac{\left( \varepsilon_F+\mathbb{E}_{+}\right) ^{3/2}}{2E}-\frac{3}{2}
\sqrt{\varepsilon_F+\mathbb{E}_{+}}\right] & \text{for }E-\varepsilon_F>\mathbb{E}_{+} \\ 
- \left [ A^{vol}_{-} \pm \eta^{vol} \right ] \frac{(\varepsilon_F-E-\mathbb{E}_{-})^2}{(\varepsilon_F-E-\mathbb{E}_{-})^2+(\mathbb{E}_{-})^2} & \text{for }E-\varepsilon_{F}<-\mathbb{E}_{-} \\ 
0 & \text{otherwise}.
\end{cases} 
\label{eq:Wnmnl}
\end{equation} 
\end{widetext}

To describe the energy dependence of surface absorption we employed the form of Ref.~\cite{Charity:2007}.
\begin{eqnarray}
W^{sur}_{\pm}\left( E\right) =\omega _{4}(E,A^{sur},B^{sur_1},0)- \nonumber \\
\omega_{2}(E,A^{sur},B^{sur_2},C^{sur}),  \label{eq:paranl} 
\end{eqnarray}
where
\begin{eqnarray}
\omega _{n}(E,A^{sur},B^{sur},C^{sur})=A^{sur}\;\Theta \left(
X\right) \frac{X^{n}}{X^{n}+\left( B^{sur}\right) ^{n}}, \nonumber \\
\label{eq:omega}
\end{eqnarray}%
and $\Theta \left( X\right) $ is Heaviside's step function and $%
X=\left\vert E-\varepsilon_F\right\vert -C^{sur}$. 
The imaginary spin-orbit term in Eq.~\eqref{eq:imnl} has the same form as the real spin-orbit
term in Eq.~\eqref{eq:HFso},
\begin{align}
      W_{so}(\bm{r},\bm{r'};E)= \left( \frac{\hbar}{m_{\pi }c}\right) ^{2}
      W^{so}(E)\frac{1}{\tilde{r}}\frac{d}{d\tilde{r}}f(\tilde{r},r^{so}_{(p,n)},a^{so})\nonumber \\
      \times\bm{\ell}\cdot \bm{\sigma}
      H(\bm{s};\beta^{so}), 
   \label{eq:imag_so}
\end{align}
where the radial parameters for the imaginary component are the same as those used for the real part of the spin-orbit potential. 
It is important to note that $\textrm{Im}\Sigma_{so}$ grows with increasing $\ell$, and for large $\ell$ this can lead to an inversion of the sign of the self-energy, which results in negative occupation. While the
form of Eq.~\eqref{eq:HFso} suppresses this behavior, it is still not a proper solution. One must be careful that the magnitude of $W_{so}(E)$ does not exceed that of the volume and surface
components. As the imaginary spin-orbit component is
generally needed only at high energies, the form of Ref.~\cite{Mueller:2011} is employed, 
\begin{equation}
   W^{so}(E)= A_{sym}^{so}  \frac{(E-\varepsilon_F)^4}{(E-\varepsilon_F)^4+(B_{sym}^{so})^4} .
   \label{eq:ImSO}
\end{equation}%

All ingredients of the self energy have now been identified and their functional form described.
In addition to the Hartree-Fock contribution and the absorptive potentials, we also include the dispersive real part from all imaginary contributions according to the corresponding subtracted dispersion relation (see Eq.~(\ref{eq:dispersion})).

\subsection*{Parameters}

 Table~\ref{table:c12} displays the parameters for the $^{12}$C self-energy.
The constraint of the number of particles was incorporated to include contributions from $\ell = 0$ to 10.
Such a range of $\ell$-values generates a sensible convergence with $\ell$ when short-range correlations are included as in Ref.~\cite{Dussan:2011}. 
We obtain 6.1 protons from all $\ell = 0$ to 10 partial wave terms including $j = \ell \pm \frac{1}{2}$ and 6.1 for neutrons. The corresponding binding energy can be found in the main text.

\subsection*{Fit Results}

\begin{figure}[h]
         \includegraphics{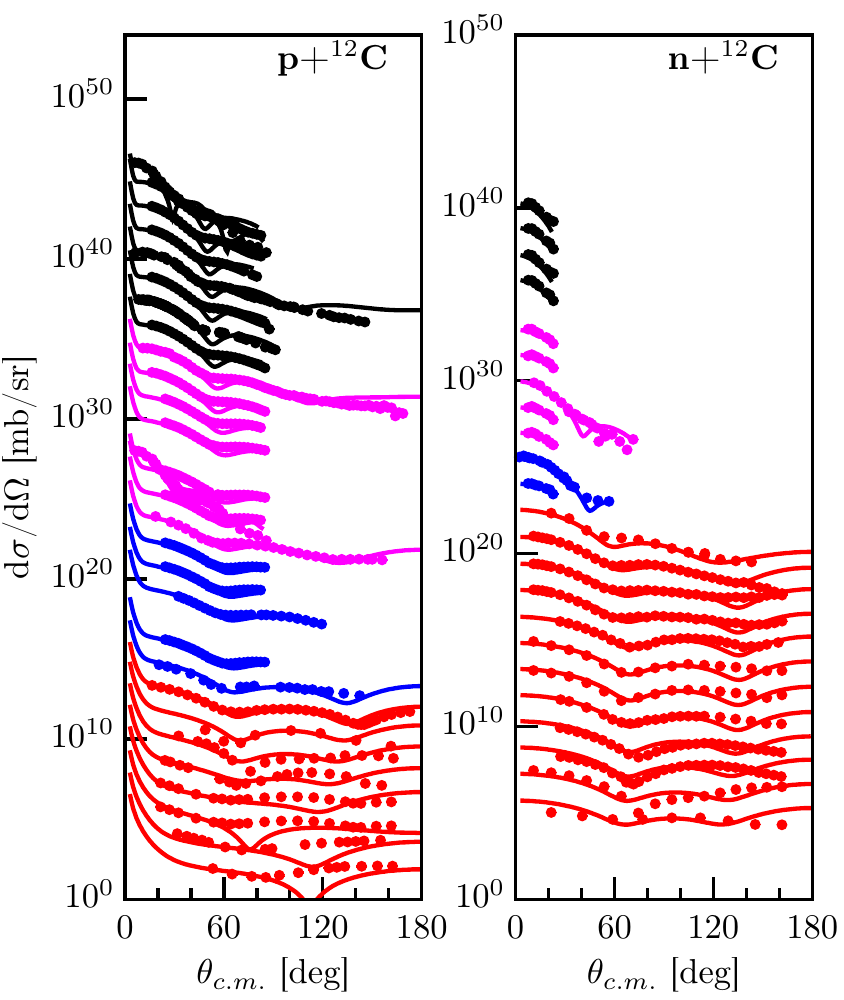}
   \caption{Calculated and experimental proton and neutron elastic-scattering angular distributions of the differential cross section $\frac{d\sigma}{d\Omega}$ at energies ranging up to 200 MeV. The data at each energy is offset by factors of ten to help visualize all of 
   the data at once. Refs.~\cite{Lebedev:2006,Wienhard:1972,Dayton:1956,Swint:1966,Kobayashi:1970,Girod:1970,Ieri:1987,An:2003,Dickens:1963,Blumberg:1966,Grantsev:1983,Rush:1971,Comfort:1980,Emmerson:1966,Hannen:2003,Comparat:1974,Meyer:1983,Johansson:1961,Ingemarsson:1979} contain the proton experimental data. Refs.~\cite{Ibaraki:2002,White:1980,Haouat:1978,Glasgow:1976,Olsson:1989,Osborne:2004,Klug:2003} contain the neutron experimental data.}
   \label{fig:elastic}
\end{figure}

\begin{figure}[h]
         \includegraphics{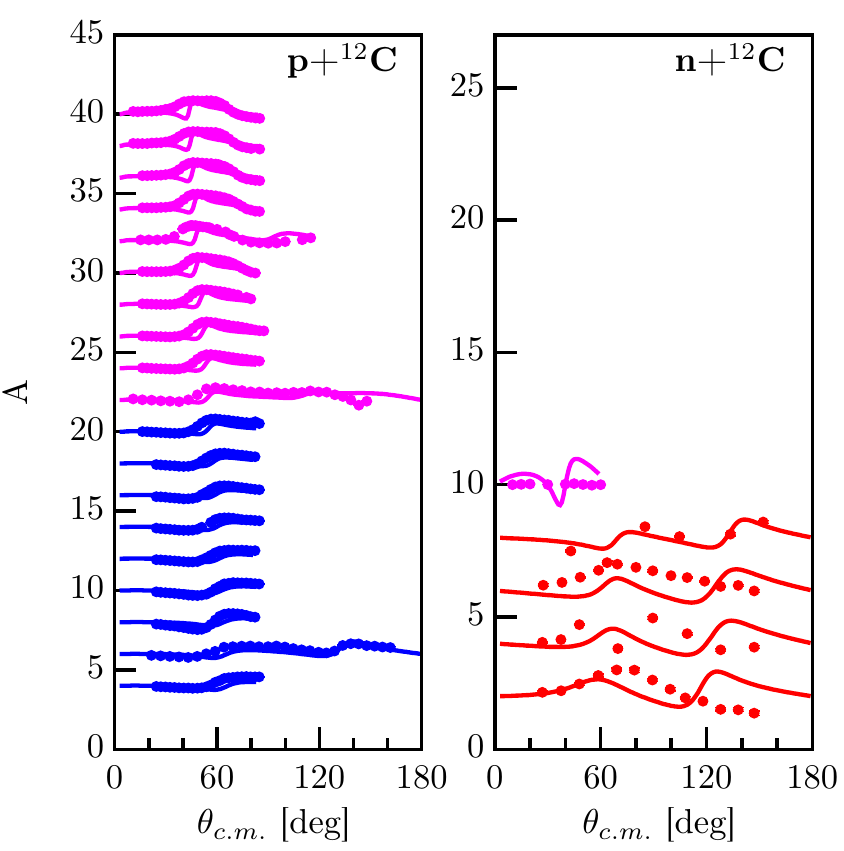}
   \caption{Calculated and experimental proton and neutron analyzing powers at energies ranging up to 200 MeV. Refs.~\cite{Sydow:1993,Ieri:1987,Craig:1966,Blumberg:1966,Kato:1980,Bauhoff:1983,Hannen:2003} contain the proton experimental data. Refs.~\cite{Roper:2005,White:1980,Hillman:1956} contain the neutron experimental data.}
   \label{fig:anal}
\end{figure}

\begin{figure}[tbp]
         \includegraphics{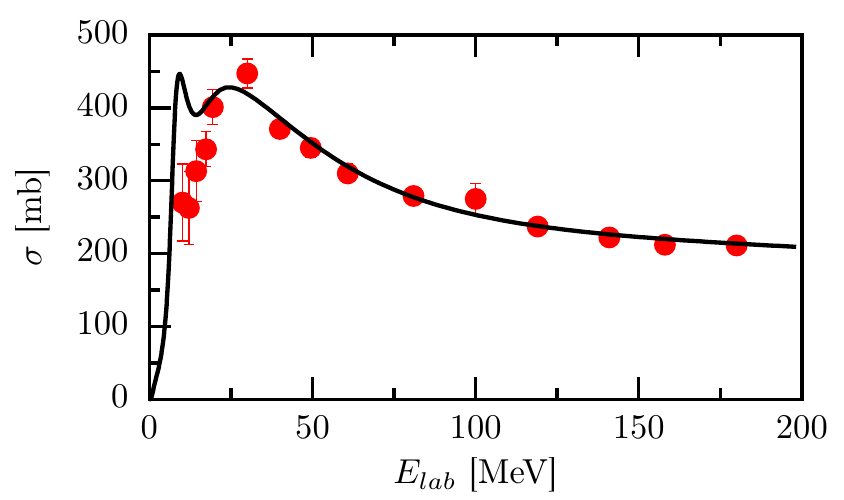}
   \caption{Proton reaction cross section generated from the DOM self-energy. The experimental data can be found in Refs.~\cite{Dicello:1970,Menet:1971,Auce:2005}.} 
   \label{fig:react}
\end{figure}

\begin{figure}[h]
         \includegraphics{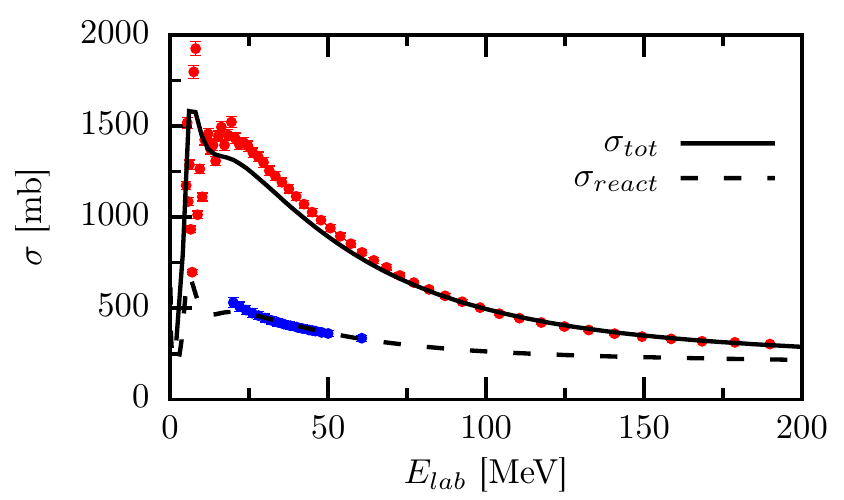}
   \caption{Neutron total cross section (solid line) and reaction cross section (dashed line) generated from the DOM self-energy. The total cross section data can be found in Ref.~\cite{Abfalterer:2001}. The reaction cross section data can be found in Ref.~\cite{Dimbylow:1980}.} 
   \label{fig:tot}
\end{figure}

\begin{figure}[h]
         \includegraphics{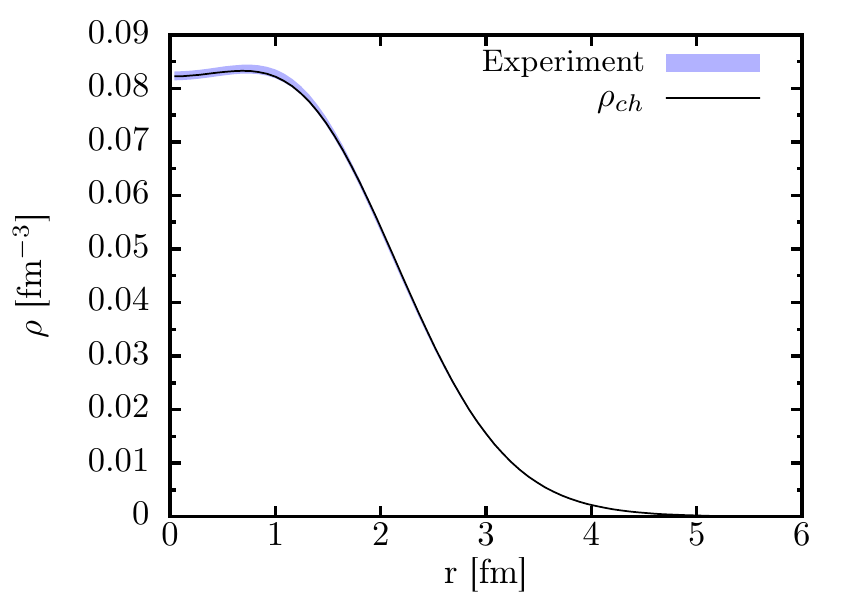}
    \caption{Experimental
       and fitted $^{12}$C charge density. The solid black line is calculated
       using the DOM self-energy and folding with the proton
       charge distribution while the experimental band represents the 1\%
       error associated with the extracted charge density from elastic
       electron scattering experiments using the sum of Gaussians
       parametrization~\cite{deVries:1987,Sick79}. 
    }
   \label{fig:chd}
\end{figure}

We found the DOM self-energy by minimizing the $\chi^2$ using experimental data in the form of elastic-scattering cross sections, total and reaction cross sections, charge density, and particle number. 
The resulting elastic-scattering cross sections are shown in Fig.~\ref{fig:elastic}, the proton analyzing powers are shown in Fig.~\ref{fig:anal}, the proton reaction cross section is shown in Fig.~\ref{fig:react}, and the neutron total cross section is shown in Fig.~\ref{fig:tot}. The charge density is shown in Fig.~\ref{fig:chd}.

\begin{table}[h]
\caption{Fitted parameter values for proton and neutron potentials in 
$^{12}$C.}
\label{table:c12}
\begin{ruledtabular}
\begin{tabular}{cc}
Parameter &                                  Value  \\
\hline
\multicolumn{2}{c}{Hartree-Fock} \\
\hline
$V^{HF}$ [MeV]		& 90.8 \\
$r^{HF}$ [fm]                       &  0.952   \\
$a^{HF}$ [fm]                       &  0.417   \\
$\beta^{vol_1}$ [fm]       &  0.908     \\
$\beta^{vol_2}$ [fm]       &  0.738     \\
$x$ 				& 0.911	\\
$\rho^{wb}$ [fm]                  &  1.01   \\
$\beta^{wb}$ [fm]                 &  0.0251   \\
\hline
\multicolumn{2}{c}{Spin-orbit} \\
\hline
$V^{so} [MeV]$                    & 26.6    \\
$r^{so}$ [fm]                       &  0.540    \\
$a^{so}$ [fm]                       &  0.755    \\
$\beta^{so}$ [fm]       &  1.23   \\
$A^{so}$ [MeV]          &      -1.62     \\
$B^{so}$ [MeV]          &      66.4        \\
\hline
\multicolumn{2}{c}{Volume imaginary} \\
\hline
$a^{vol}_{+}$ [fm] 		& 0.536 	\\ 
$r^{vol}_{+}$ [fm] 		& 1.27 	\\ 
$\beta^{vol}_{+}$ [fm]		& 0.340	\\ 
$a^{vol}_{-}$ [fm] 		& 0.256 	\\ 
$r^{vol}_{-}$ [fm] 		& 1.02 	\\ 
$\beta^{vol}_{-}$	[fm]		& 1.08 \\
$A^{vol}_{+}$ [MeV]           & 6.51  \\
$B^{vol}_{+}$ [MeV]		& 25.3	\\
$\mathcal{E}^{vol}_{+}$ [MeV]	& 2.31	\\
$A^{vol}_{-}$ [MeV]            & 16.9  \\
$B^{vol}_{-}$ [MeV]		& 8.97	\\
$\mathcal{E}^{vol}_{-}$ [MeV]		& 1.61	 \\
$\mathbb{E}_{+}$ [MeV]			& 23.4	 \\
$\mathbb{E}_{-}$ [MeV]			& 67.5	 \\
$\alpha$                      & 0.189  \\
\hline
\multicolumn{2}{c}{Surface imaginary} \\
\hline
$a^{sur}_{+}$ [fm] 		& 0.493 	 \\ 
$r^{sur}_{+}$ [fm] 		& 1.40 	 \\ 
$\beta^{sur}_{+}$ [fm]       & 3.38   \\ 
$a^{sur}_{-}$ [fm] 		& 0.316 	 \\ 
$r^{sur}_{-}$ [fm] 		& 0.631 	 \\ 
$\beta^{sur}_{-}$ [fm]       & 1.72   \\ 
$A^{sur}_{+}$ [MeV]          & 13.0  \\
$B^{sur_1}_{+}$ [MeV]         & 26.3  \\
$B^{sur_2}_{+}$ [MeV]         & 198  \\
$C^{sur}_{+}$ [MeV] & 199    \\
$A^{sur}_{-}$ [MeV]          & 28.0   \\
$B^{sur_1}_{-}$ [MeV]         & 23.11  \\
$B^{sur_2}_{-}$ [MeV]         & 20.0  \\
$C^{sur}_{-}$ [MeV] & 94.9  \\
\end{tabular}
\end{ruledtabular}
\end{table}

\section*{References}
\label{sec:refs}
\bibliographystyle{apsrev4-1}
\bibliography{energy}

\end{document}